\documentclass{ws-procs9x6}

\begin{document}

\title{Observation of Galactic Sources of Very High Energy $\gamma$-Rays with the MAGIC Telescope}

\author{H. Bartko (for the MAGIC collaboration)}

\address{MPI f\"ur Physik, Werner-Heisenberg-Institut,\\
F\"ohringer Ring 6, D-80805 M\"unchen, Germany\\
E-mail: hbartko@mppmu.mpg.de}



\begin{abstract}
During its first cycle of observations, the MAGIC (Major Atmospheric Gamma-ray Imaging Cherenkov) telescope has observed very high energy $\gamma$-rays from five galactic objects: the Crab Nebula, the SNRs HESS J1813-178 and HESS J1834-087, the Galactic Center and the $\gamma$-ray binary LS~I~+61~303. After a short introduction to the MAGIC telescope and the data analysis procedure, the results of these five sources are reviewed.
\end{abstract}

\keywords{$\gamma$-ray astronomy, Galactic objects, Galactic Center.}

\bodymatter

\section{The MAGIC Telescope} 
\label{sec:intro}
MAGIC \cite{MAGIC-commissioning,CortinaICRC} 
is currently the largest single dish Imaging
Air Cherenkov Telescope (IACT) in operation. Located on the Canary
Island La Palma ($28.8^\circ$N, $17.8^\circ$W, 2200~m a.s.l.), it 
has a 17-m diameter tessellated parabolic mirror,
supported by a light weight carbon fiber frame. It is equipped
with a high quantum efficiency 576-pixel $3.5^\circ$ field-of-view photomultiplier
camera. The analog signals are transported via optical fibers to
the trigger electronics
and the 300 MSamples/s FADC read-out system, which is currently being upgraded to a 2 GSamples/s FADC system \cite{MUX_tests}. 


The physics program of the MAGIC telescope includes both,
topics of fundamental physics and astrophysics. In this paper 
the observations of galactic sources are presented. 
The observations of extragalactic sources are reviewed
elsewhere in these proceedings~\cite{errando,garcz}.

\section{Data Analysis}
\label{sec:data_analysis}

The data analysis was carried out using the standard MAGIC
analysis and reconstruction software \cite{Magic-software}, the
first step of which involves the calibration of the raw data
\cite{MAGIC_calibration,OF_Magic}. It follows the general steps presented
in \cite{MAGIC_1813,MAGIC_GC}: After calibration, image cleaning
tail cuts have been applied (see e.g. \cite{Fegan1997}).
The camera images are parameterized by
image parameters \cite{Hillas_parameters}. The
Random Forest method (see \cite{RF,Breiman2001} for a detailed
description) was applied for the $\gamma$/hadron separation (for a
review see e.g. \cite{Fegan1997}) and the energy estimation.



For each event the arrival direction of the primary $\gamma$-ray
candidate in sky coordinates is estimated using the DISP-method resulting in VHE $\gamma$-ray sky map
\cite{wobble,Lessard2001,MAGIC_disp}. The angular resolution is $\sim 0.1^\circ$, while source localization in the sky is provided with a precision of $\sim 2'$ \cite{starguider}.

The points of the reconstructed
very high energy $\gamma$-ray spectrum
($\mathrm{dN}_{\gamma}/(\mathrm{dE}_{\gamma} \mathrm{dA}
\mathrm{dt})$ vs. true $\mathrm{E}_{\gamma}$) are 
corrected (unfolded) for the instrumental energy resolution
\cite{Anykeev1991}. 
Moreover, a forward unfolding 
procedure is applied for spectral fits: A candidate spectral law is fitted to the measured
data ($\mathrm{dN}_{\gamma}/(\mathrm{dE}_{\gamma} \mathrm{dA}
\mathrm{dt})$ vs. estimated $\mathrm{E}_{\gamma}$) taking the full
instrumental energy migration (true $\mathrm{E}_{\gamma}$ vs.
estimated $\mathrm{E}_{\gamma}$) into account as described in
\cite{Mizobuchi2005}. 


The systematic error in the flux level determination is
estimated to be 35\% 
the spectral index 0.2, see also \cite{MAGIC_GC}.


\section{Highlights of cycle I} 
\label{sec:highlights}

MAGIC's first observation cycle spanned the period from January 2005
to April 2006. About 1/4 of the scientific observation time 
was devoted to galactic objects. The observations covered 
included the following
types of objects: superova remnants (SNRs), pulsars, pulsar wind
nebulae (PWN), microquasar candidates ($\mu$QSRs), the Galactic Center (GC), one
unidentified TeV source and one cataclysmic variable. In this section the results of the following sources are reviewed: the Crab nebula and some other selected pulsars, the SNRs HESS J1813-178 and HESS J1834-087, the Galactic Center and the $\gamma$-ray binary LS~I~+61~303.

\subsection{The Crab nebula and pulsars} 
\label{sec:crab}
The Crab nebula is a bright and steady emitter at GeV and TeV energies, what
makes it into an excellent calibration candle. This object has
been observed extensively in the past over a wide range of
wavelengths. Some of the relevant physics phenomena,  
are expected to happen in the energy domain between 10 and 100~GeV,
namely the Inverse Compton peak of the energy sectral energy distribution 
and the cut-off of the pulsed emission.

\begin{figure}[h]
  \centering
  \includegraphics[width=7.5cm]{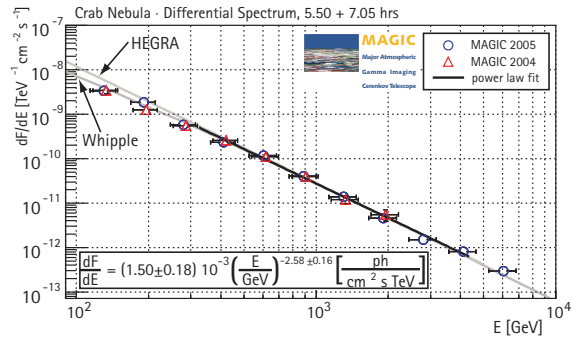}
  \caption{ Energy spectrum above 100 GeV from the Crab nebula
  measured by MAGIC in two different observation seasons~\cite{Crab_MAGIC}.}
  \label{fig:crab}
\end{figure}

Along the first cycle of MAGIC's regular observations, a significant
amount of time has been devoted to observe the Crab nebula, both for
technical and astrophysical studies. A sample of
12~hours of selected data has been used to measure
the energy spectrum down to $\sim$100~GeV, as shown in
figure~\ref{fig:crab}~\cite{Crab_MAGIC}. Also a search
for pulsed $\gamma$-ray emission from the Crab pulsar has been carried out. The derived
upper limits (95\% C.L.) are 2.0$\times 10^{-10}$~ph s$^{-1}$cm$^{-2}$ at 90~GeV and
1.1$\times 10^{-10}$~ph s$^{-1}$cm$^{-2}$ at 150~GeV \cite{Crab_pulsar_MAGIC}.

Moreover, $\gamma$-ray emission was searched for from two
milisecond pulsars~\cite{ona} PSR~B1957+20 and PSR~J0218+4232,
albeit without positive result. The corresponding upper limits ($E_{\gamma} \geq 115$~GeV) are
$F_{\mathrm{PSR~B1957+20}} \sim 2.3 \times 10^{-11}$ and 
$F_{\mathrm{PSR~J0218+4232}} \sim 2.9 \times 10^{-11}$ ph s$^{-1}$cm$^{-2}$ for
the steady emission and  
$F_{\mathrm{PSR~J0218+4232}} \sim 6.5 \times 10^{-12}$ ph s$^{-1}$cm$^{-2}$ for
the pulsed one.


\vspace{-0.3cm}
\subsection{Supernova remnants}
\label{sec:sn}


Shocks produced at supernova explosions are assumed to be the source
of the galactic component of the cosmic ray flux~\cite{zwicky}. 
In inelastic collisions of high energy cosmic rays with ambient matter $\gamma$-rays and neutrinos are produced. These neutral particles give direct information about their source, as their trajectories are not affected by magnetic fields in contrast to the charged cosmic rays. Nevertheless, not all VHE $\gamma$-rays from galactic sources are due to the interactions of cosmic rays with ambient matter. There are also other mechanisms for the production of VHE $\gamma$-rays like the inverse Compton up-scattering of ambient low energy photons by VHE electrons. For each individual source of VHE $\gamma$-rays, the physical processes of particle acceleration and $\gamma$-ray emission in this source have to be determined. A powerfull tool is the modelling of the multiwavelength emission of the source and comparison to multiwavelength data.

Within its program of observation of galactic sources, MAGIC has
taken data on a number of supernova remnants. 
VHE $\gamma$-ray emission from the SNRs
HESS~J1813-178~\cite{MAGIC_1813} and HESS~J1834-087
(W41)~\cite{MAGIC_1834} has been observed. These results have confirmed SNRs as a well
established population of VHE $\gamma$-ray emitters. 

The energy spectrum of HESS~J1813-178, which is spatially coincident with SNR G12.82-0.02, can be fitted with a hard-slope power law, described as $\mathrm{d}N_{\gamma}/(\mathrm{d}A \mathrm{d}t \mathrm{d}E) = (3.3 \pm 0.5) \times 10^{-12} (E/\mathrm{TeV})^{-2.1 \pm 0.2} \ \mathrm{cm}^{-2}\mathrm{s}^{-1} \mathrm{TeV}^{-1}$. The source HESS J1834-087 is spatially coincident with SNR G23.3-0.3 (W41). The observed differential $\gamma$-ray flux is consistent with a power law $\mathrm{d}N_{\gamma}/(\mathrm{d}A
\mathrm{d}t \mathrm{d}E) = (3.7 \pm 0.6) \times 10^{-12}
 (E/\mathrm{TeV})^{-2.5 \pm 0.2} \ \mathrm{cm}^{-2}\mathrm{s}^{-1}
\mathrm{TeV}^{-1}$. A source extension of $(0.14\pm0.04)^\circ$ is derived. A spatial superposition of the $\gamma$-ray source with a massive molecular cloud observed by its $^{13}$CO and $^{12}$CO emission was found. Although the mechanism responsible
for the VHE radiation remains yet to be clarified, this is a hint that
it could be produced by high energy hadrons interacting with the
molecular cloud.

\subsection{Galactic Center}
\label{sec:gc}

The Galactic Center region contains many remarkable objects which may be responsible for high-energy processes generating $\gamma$-rays like a super-massive black hole, supernova remnants, candidate pulsar wind nebulae, a high density of cosmic rays, hot gas and large magnetic fields. Moreover, the Galactic Center may emit the highest VHE $\gamma$-ray flux from the annihilation of possible dark matter particles \cite{DM_ICRC} of all proposed dark matter particle annihilation sources. 

One motivation for this observations was the possibility to indirectly detect dark matter
through its annihilation into VHE $\gamma$-rays, see e.g. \cite{DM_ICRC}. 

The Galactic Center was observed with the MAGIC telescope~\cite{MAGIC_GC} under large zenith angles resulting in the detection of a differential $\gamma$-ray flux, consistent with a steady, hard-slope power law between 500~GeV and about 20~TeV, described as $\mathrm{d}N_{\gamma}/(\mathrm{d}A \mathrm{d}t \mathrm{d}E) = (2.9 \pm 0.6)\times 10^{-12} (E/\mathrm{TeV})^{-2.2 \pm 0.2} \ \mathrm{cm}^{-2}\mathrm{s}^{-1} \mathrm{TeV}^{-1}$. This result confirms the previous measurements by the HESS collaboration. The VHE $\gamma$-ray emission does not show any significant time variability; the MAGIC measurements rather affirm a steady emission of $\gamma$-rays from the GC region on time scales of up to one year.

The VHE $\gamma$-ray source is centered at (RA, Dec)=(17$^{\mathrm{h}}45^{\mathrm{m}}20^{\mathrm{s}}$, -29$^\circ2'$). The excess is point-like, it's location is consistent with SgrA$^*$, the candidate PWN G359.95-0.04 as well as SgrA East.

The nature of the source of the VHE $\gamma$-rays has not yet been identified. The power law spectrum up to about 20~TeV disfavours dark matter annihilation as the main origin of the detected flux.
The absence of flux variation indicates that the VHE $\gamma$-rays are rather produced in a steady object such as a SNR or a PWN, and not in the central black hole.



\subsection{The $\gamma$-ray binary LS~I~+61~303} 
\label{sec:lsi}

\begin{figure}[!t]
\centering
\includegraphics[width=\textwidth]{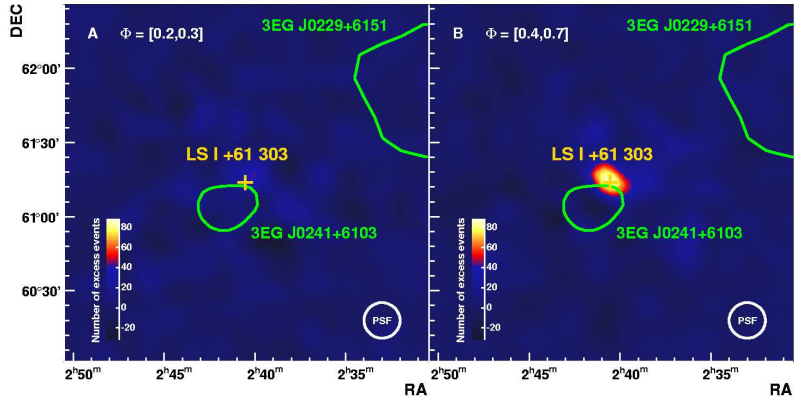}
\caption{
Smoothed maps of $\gamma$-ray excess events above 400 GeV around
LS~I~+61~303, for observations around periastron (A) and latter orbital phases (B)~\cite{lsi}.
}
\label{fig:lsi-skymap}
\end{figure}

This $\gamma$-ray binary system is composed of a B0 main sequence star
with a circumstellar disc, i.e. a Be star, located at a distance of
$\sim$2 kpc. A compact object of unknown nature (neutron star or black
hole) is orbiting around it, in a highly eccentric ($e=0.72\pm0.15$)
orbit.
  
LS~I~+61~303 was observed with MAGIC for 54 hours
between October 2005 and March 2006~\cite{lsi}. 
The reconstructed $\gamma$-ray map is shown in
figure~\ref{fig:lsi-skymap}. The data were first divided into 
two different samples, around periastron passage (0.2-0.3)
and at higher (0.4-0.7) orbital phases. No significant excess in the
number of $\gamma$-ray events is detected around periastron passage,
whereas there is a clear detection (9.4$\sigma$ statistical significance) at
later orbital phases.
Two different scenarios have been involved to explain this high energy
emissions: the microquasar scenario where the $\gamma$-rays are produced
in a radio-emitting jet; or the pulsar binary scenario, where 
they are produced in the shock which is generated by the interaction
of a pulsar wind and the wind of the massive companion.
See~\cite{sidro} for more details.

\paragraph{Acknowledgements.}
We thank the IAC for the excellent working conditions at the
ORM in La Palma. The support of the
German BMBF and MPG, the Italian INFN, the Spanish CICYT is gratefully
acknowledged. This work was also supported by ETH research grant
TH-34/04-3, and the Polish MNiI grant 1P03D01028. 

\vspace{-0.5cm}

\bibliographystyle{ws-procs9x6}

\end{document}